\documentclass[manuscript]{aastex}

\usepackage[utf8]{inputenc}
\usepackage{lmodern}
\usepackage{amsmath}
\usepackage{amssymb}
\usepackage{mathrsfs}

\usepackage{color}				
\definecolor{red}{rgb}{0,0,0}		
\definecolor{blue}{rgb}{0,0,0}		
\definecolor{black}{rgb}{0,0,0}		
\newcommand\red{\color{red}}
\newcommand\blu{\color{blue}}
\newcommand\bla{\color{black}}

\slugcomment{Submitted to ApJ Letters}

\shorttitle{Pluto's atmosphere in 2015}
\shortauthors{Occultation's Team et al.}

\begin{document}


\title{%
Pluto's atmosphere from the  29 June 2015 ground-based stellar occultation at the time of the New Horizons 
flyby\footnote{%
Partly based on observations made 
with the ESO WFI camera at the 2.2 m Telescope (La Silla), under program ID 079.A-9202(A) 
within the agreement between the ON/MCTI and the Max Planck Society, 
with the ESO camera NACO at the Very Large Telescope (Paranal), under program ID 089.C-0314(C), and
at the Pico dos Dias Observatory/LNA, Brazil.
}%
}%


\author{%
B. Sicardy\altaffilmark{1},
J. Talbot\altaffilmark{2},
E. Meza\altaffilmark{1},
J.~I.~B. Camargo\altaffilmark{3,4},
J. Desmars\altaffilmark{5},
D. Gault\altaffilmark{6,7},
D. Herald\altaffilmark{2,7,8},
S. Kerr\altaffilmark{2,9},
H. Pavlov\altaffilmark{7},
F. Braga-Ribas\altaffilmark{10,3,4},
M. Assafin\altaffilmark{11},
G. Benedetti-Rossi\altaffilmark{3},
A. Dias-Oliveira\altaffilmark{3},
A. R. Gomes-J\'unior\altaffilmark{11},
R. Vieira-Martins\altaffilmark{3},
D. B\'erard\altaffilmark{1},
P. Kervella\altaffilmark{1,12},
J. Lecacheux\altaffilmark{1},
E. Lellouch\altaffilmark{1},
W. Beisker\altaffilmark{13},
D. Dunham\altaffilmark{7},
M. Jel\'{\i}nek\altaffilmark{14,15},
R. Duffard\altaffilmark{15},
J.~L. Ortiz\altaffilmark{15},
A.~J. Castro-Tirado\altaffilmark{15,16},
R. Cunniffe\altaffilmark{15},
R. Querel\altaffilmark{17},
P.~C. Yock\altaffilmark{18},
A.~A. Cole\altaffilmark{19},
A.~B. Giles\altaffilmark{19}, 
K.~M. Hill\altaffilmark{19},
J.~P. Beaulieu\altaffilmark{20},
M. Harnisch\altaffilmark{2,21},
R. Jansen\altaffilmark{2,21},
A. Pennell\altaffilmark{2,21},
S. Todd\altaffilmark{2,21},
W.~H. Allen\altaffilmark{2},
P.~B. Graham\altaffilmark{2,22},
B. Loader\altaffilmark{7,2},
G. McKay\altaffilmark{2},
J. Milner\altaffilmark{2},
S. Parker\altaffilmark{23},
M.~A. Barry\altaffilmark{2,24},
J. Bradshaw\altaffilmark{7,25},
J. Broughton\altaffilmark{2},
L. Davis\altaffilmark{6},
H. Devillepoix\altaffilmark{26},
J. Drummond\altaffilmark{27},
L. Field\altaffilmark{2},
M. Forbes\altaffilmark{2,22},
D. Giles\altaffilmark{6,28},
R. Glassey\altaffilmark{29},
R. Groom\altaffilmark{30},
D. Hooper\altaffilmark{2},
R. Horvat\altaffilmark{6},
G. Hudson\altaffilmark{2},
R. Idaczyk\altaffilmark{2},
D. Jenke\altaffilmark{31},
B. Lade\altaffilmark{31},
J. Newman\altaffilmark{8},
P. Nosworthy\altaffilmark{6},
P. Purcell\altaffilmark{8},
P.~F. Skilton\altaffilmark{2,32},
M. Streamer\altaffilmark{8},
M. Unwin\altaffilmark{2},
H. Watanabe\altaffilmark{33},
G. L. White\altaffilmark{6},
\and
D. Watson\altaffilmark{2}
%
%
}
\email{bruno.sicardy@obspm.fr}

\altaffiltext{1}{LESIA/Observatoire de Paris, PSL, CNRS UMR 8109, University Pierre et Marie Curie, 
University Paris-Diderot, 5 place Jules Janssen, F-92195 Meudon C\'edex, France}

\altaffiltext{2}{Occultation Section of the Royal Astronomical Society of New Zealand (RASNZ)}

\altaffiltext{3}{Observat\'orio Nacional/MCTI, R. General Jos\'e Cristino 77, Rio de Janeiro - RJ, 20.921-400, Brazil}

\altaffiltext{4}{Laborat\'orio Interinstitucional de e-Astronomia - LIneA, Rua Gal. Jos\'e Cristino 77, Rio de Janeiro, RJ 20921-400, Brazil}

\altaffiltext{5}{IMCCE/Observatoire de Paris, 77 Avenue Denfert Rochereau, Paris, F-75014, France}

\altaffiltext{6}{Western Sydney Amateur Astronomy Group (WSAAG),  Sydney, NSW, Australia}

\altaffiltext{7}{International Occultation Timing Association (IOTA)}

\altaffiltext{8}{Canberra Astronomical Society, Canberra, ACT, Australia}

\altaffiltext{9}{Astronomical Association of Queensland, QLD, Australia}

\altaffiltext{10}{Federal University of Technology - Paran\'a (UTFPR/DAFIS), 
Rua Sete de Setembro, 3165, CEP 80230-901, Curitiba, PR, Brazil}

\altaffiltext{11}{Universidade Federal do Rio de Janeiro,  Observat\'orio do Valongo, 
Ladeira do Pedro Ant\^onio 43, CEP 20080-090, Rio de Janeiro, Brazil}

\altaffiltext{12}{Unidad Mixta Internacional Franco-Chilena de Astronom\'{i}a (UMI 3386), CNRS/INSU, France \&
Departamento de Astronom\'{i}a, Universidad de Chile, Camino El Observatorio 1515, Las Condes, Santiago, Chile}

\altaffiltext{13}{IOTA/ES, Bartold-Knaust-Strasse 8, D-30459 Hannover, Germany}

\altaffiltext{14}{Astronomical Institute of the Czech Academy of Sciences, Fri\v{c}ova 298, CZ-25165 Ond\v{r}ejov,
Czech Republic}

\altaffiltext{15}{Instituto de Astrof\'isica de Andaluc\'ia-CSIC, Aptd 3004, E-18080, Granada, Spain}

\altaffiltext{16}{\blu 
%
Departamento  de  Ingenier\'ia de  Sistemas  y  Autom\'atica,  
E.T.S.  de Ingenieros Industriales, Universidad de M\'alaga, 
Unidad Asociada al CSIC, M\'alaga, Spain
}

\altaffiltext{17}{National Institute of Water and Atmospheric Research (NIWA), Lauder, New Zealand}

\altaffiltext{18}{Department of Physics, University of Auckland, Private Bag 92019, Auckland, New Zealand}

\altaffiltext{19}{School of Physical Sciences, University of Tasmania, Private Bag 37, Hobart, TAS 7001, Australia}

\altaffiltext{20}{Sorbonne Universit\'es, Universit\'e Pierre et Marie Curie et CNRS, UMR 7095, 
Institut d'Astrophysique de Paris, 98 bis bd Arago, 75014 Paris, France}

\altaffiltext{21}{Dunedin Astronomical Society, Dunedin, New Zealand}

\altaffiltext{22}{Wellington Astronomical Society (WAS), Wellington, New Zealand}

\altaffiltext{23}{Backyard Observatory Supernova Search (BOSS), Australia and New Zealand}

\altaffiltext{24}{ Univ. of Sydney, Elect. and Info. Engineering Dpt, Camperdown, 2006 NSW, Australia}

\altaffiltext{25}{Samford Valley Observatory, QLD, Australia}

\altaffiltext{26}{International Centre for Radio Astronomy Research (ICRAR), and the Department of Applied Geology, 
Curtin University, Bentley, WA 6102, Australia}

\altaffiltext{27}{Possum Observatory, Patutahi, New Zealand}

\altaffiltext{28}{Penrith Observatory, Western Sydney University, Sydney,  NSW, Australia}

\altaffiltext{29}{Canterbury Astronomical Society, Christchurch, New Zealand}

\altaffiltext{30}{Astronomical Society of Western Australia, P.O. Box 421, Subiaco, Perth, WA 6904, Australia}

\altaffiltext{31}{Stockport Observatory, Astronomical Society of South Australia, Stockport, SA, Australia}

\altaffiltext{32}{Mornington Peninsula Astronomical Society, Mount Martha, VIC, Australia}


\altaffiltext{33}{Japan Occultation Information Network (JOIN), Japan}


\begin{abstract}
We present results from a multi-chord Pluto stellar occultation observed on 29 June 2015
from New Zealand and Australia.
This occurred only two weeks before the NASA New Horizons flyby of the Pluto system and
serves as a useful comparison between ground-based and space results.
%
We find that Pluto's atmosphere is still expanding, with a significant pressure increase of 5$\pm$2\%
since 2013 and a factor of almost three since 1988.
This trend rules out, as of today, an atmospheric collapse associated with
Pluto's recession from the Sun. 
A central flash, a rare occurrence,  was observed from several sites in New Zealand. 
The flash shape and amplitude are compatible with a
spherical and transparent atmospheric layer of roughly 3~km in thickness 
whose base lies at about 4~km above Pluto's surface, and
where an average thermal gradient of about 5 K~km$^{-1}$ prevails.
We discuss the possibility that small departures between the observed and modeled flash
are caused by local topographic features (mountains) along Pluto's limb that block the stellar light.
Finally, 
using two possible temperature profiles, and extrapolating our pressure profile
from our deepest accessible level down to the surface, 
we obtain a \red 
possible range of 11.9-13.7~$\mu$bar \bla
for the surface pressure.
\end{abstract}






\section{Introduction}

Ground-based stellar occultations probe Pluto's atmosphere 
at radii ranging from 
$r \sim1190$~km from the planet center (pressure $p \sim10$~$\mu$bar) up to 
$r \sim 1450$~km ($p \sim 0.1$~$\mu$bar).
In a previous work \citep[DO15 hereafter]{dia15},
we analyzed high signal-to-noise-ratio 
occultations observed in 2012 and 2013, and derived  stringent constraints on Pluto's 
atmospheric profiles (density, pressure and temperature profiles).
and on Pluto's radius ($R_P = 1190\pm5$~km, assuming no troposphere).
We also found a pressure increase of $6\pm1$\% between 2012 and 2013.

Here we analyze a stellar occultation, observed on 29 June 2015 from Australia and New Zealand, 
which occurred two weeks before the NASA New Horizons (NH hereafter) flyby of the Pluto system.
Our goals are: 
(1) assess further pressure changes between 2013 and 2015
(eventually providing useful constraints on Pluto's seasonal models); 
(2) analyze the central flash that was detected for the first time ever from multiple stations.
It  constrains the thermal structure of a layer immediately above Pluto's surface, 
its possible departure from sphericity and/or presence of hazes; 
%
and (3) constrain the pressure at Pluto's surface.  
%
Besides serving as a useful comparison with the NH results, our work
is one more benchmark in the long-term survey of Pluto's atmosphere
over the forthcoming years.

\section{The 29 June 2015 occultation}

\begin{figure}
\epsscale{0.8}
\plotone{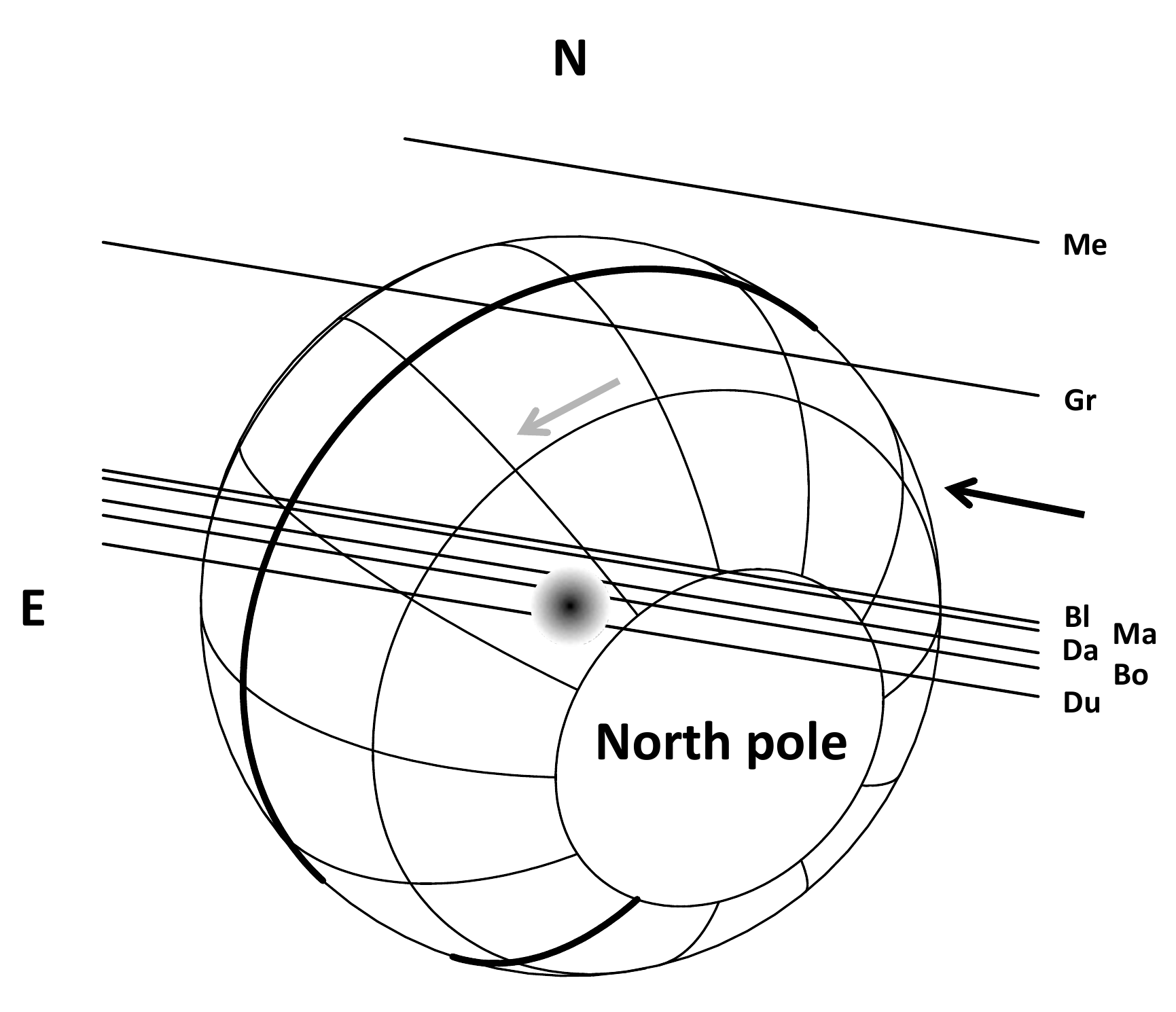}
\caption{%
Geometry of the 29 June 2015 Pluto stellar occultation.
The stellar motion relative to Pluto (black arrow) 
is  shown for seven stations,
\textbf{Me}: Melbourne,
\textbf{Gr}: Greenhill, 
\textbf{Bl}: Blenheim,
\textbf{Ma}: Martinborough,
\textbf{Da}: Darfield,
\textbf{Bo}: BOOTES-3,
\textbf{Du}: Dunedin.
The J2000 celestial north and east are indicated by N and E, respectively.
Pluto's radius is fixed at 1187~km.
The equator and prime meridian are drawn as thicker lines,
and direction of rotation is along the gray arrow.
The shaded region at center indicates the central flash  zone.
\label{fig_chords}
}%
\end{figure}

The prediction procedures are 
described in DO15, \cite{ass10} and \cite{ben14}.
The event was monitored from Australia and New Zealand (Table~\ref{tab_pos}),
from which we obtained eight occultation detections. 
%
The reconstructed occultation geometry is displayed in Fig.~\ref{fig_chords}, 
see also Table~\ref{tab_param}.
The light-curves were obtained from classical aperture photometry, after correction of low 
frequency variations (caused by changing sky conditions) by means of nearby reference stars, when available.
The resulting light-curves $\phi(t)$ give
the total flux from the star and Pluto's system, 
normalized to unity outside the occultation, as a function of time $t$
(Fig.~\ref{fit_all}). 
The observed flux $\phi$ can be written:
\begin{equation}
\phi=  (1- \phi_P) \cdot F_\star + \phi_P,
\label{eq_phi}
\end{equation}
where $F_\star$ is the (useful) stellar flux alone, normalized between zero and unity.
Thus, $\phi_P$ and $1-\phi_P$ are the contributions of Pluto's system and the unocculted 
stellar flux to $\phi$, respectively. 

The quantity 
$\phi_P$ is in principle measured independently
when Pluto and the occulted star are angularly resolved,  
providing $F_\star$.
It is difficult in practice and 
requires high photometric accuracy on the 
star, Pluto and nearby reference stars hours or days away from the event. 
During that time, sky and instrument conditions may vary. 
Moreover, for data taken without a filter (broadband), 
chromatic dependence of the extinction adds further systematic biases,
especially if calibrations are not made at the same airmass.

One station that went deep into Pluto's shadow (BOOTES-3, broadband, \citealt{cas12}) 
obtained calibration images 
hours before the event, as the star and Pluto were marginally resolved.
However, the overlap of the star and Pluto images prevents the 
useful
determination of  the Pluto/star ratio at the required accuracy (1\% or better).
%
Moreover the airmass variation (1.1 during calibration vs. 1.6 during the occultation)
introduces unmodeled chromatic effects due to color differences between the star and Pluto.
More images taken the following night at very high airmass (3.6) do not provide 
further constraints on $\phi_P$.

One light-curve (Dunedin) was affected by non-linearity caused by a 
so-called  ``$\gamma$ factor" \citep{poy96} that modified the pixel values
to increase the image dynamical range . 
The (supposedly) reverse transformation provides
an event that is globally not deep enough considering its duration,
%
indicating residual non-linearities.
Thus, for this station, we only used the bottom part of the light-curve (Fig.~\ref{fit_all}), 
assuming that in this range, the retrieved flux $\phi$ is an affine function of the stellar flux, 
$\phi = a \cdot F_\star + b$.

In spite of the lack of accurate measurements for $\phi_P$, 
the amplifying effect of the central flash still constrains the thermal structure of Pluto's 
deepest atmospheric layers (see Section~\ref{sec_flash}).

\section{Pressure evolution}
\label{sec_evolution}

The DO15 model uses the simplest possible hypotheses, i.e. Pluto's atmosphere 
(1) is pure nitrogen (N$_2$), 
(2) is spherically symmetric, 
(3) has a time-independent thermal structure, derived itself from the light-curves, and
(4) is transparent (haze-free).
The validity of hypotheses (1)-(3) is discussed in DO15.
Hypothesis (4) is discussed later in view of the NH results. 
Adjusting the pressure $p_0$ at a reference radius $r_0$ (for a given event)
uniquely defines the molecular density profile $n(r)$,
from which synthetic light-curves are generated and compared to the data.
%
%
%
%
%
%
%
Note that $p_0$ monitors the evolution of Pluto's atmospheric pressure as a whole.
%
In practice, most of the contribution to the fits comes from the 
half-light level ($F_\star \sim 0.5$, $r \sim 1295$~km, $p \sim 1.7$~$\mu$bar),
with a tapering off 
above $r \sim 1450$~km ($F_\star \sim 0.9$, $p \sim 0.1$~$\mu$bar) and 
below $r \sim 1205$~km ($F_\star \sim 0.1$, $p \sim 8$~$\mu$bar).
%

The parameters of our model are listed in Table~\ref{tab_param} and
our simultaneous fits are displayed in Fig.~\ref{fit_all}.
They have $\chi^2$ per degree of freedom  close to unity, indicating satisfactory fits.
Two minor modifications were introduced, relative to the DO15 model.
First,  we updated for consistency Pluto's mass factor to
$GM= 8.696 \times 10^{11}$ m$^3$ s$^{-2}$ \citep{ste15},
instead of $8.703 \times 10^{11}$ m$^3$ s$^{-2}$,
causing negligible changes at our accuracy level.
Second, we use the NH-derived Pluto  radius ($R_P=1187$~km)
as a boundary condition for the DO15 model.
%
%
%
%
%
This new value modifies (at a few percent level) the retrieved pressure at a given radius compared to DO15.
Moreover, changing $R_P$ translates vertically all the profiles near the surface
by an equivalent amount. In other words, all the quantities of interest (pressure, density, temperature)
are well defined in terms of altitude above the surface, if not in absolute radius.
%
%
%

The pressures $p_0$  at $r_0= 1215$~km and $1275$~km 
are given in Table~\ref{tab_param}.
They are useful benchmarks, respectively corresponding to the stratopause
(maximum  temperature of 110~K),  and
the half-light level layer.
%
Fig.~\ref{fig_t_p_p_r} displays the pressure evolution over 2012-2015.
\red
The formal error bars 
assume  an invariant temperature profile,
but this assumption should not affect the \textit{relative} pressure changes in 2012-2015. 
Relaxing that constraint, we can retrieve $p_0$ by inverting 
individual light-curves and testing the effects of the inversion parameters.
This yields possible biases estimated to $\pm 0.2,  \pm0.8$ and $\pm0.5$~$\mu$bar
in 2012, 2013 and 2015, respectively.
\bla
We have added for comparison occultation results 
from 1988 \citep{yel97} and 2002 \citep{sic03}. 
They stem from different analyses and may also be affected by  biases. 
However, Fig.~\ref{fig_t_p_p_r} should capture the main trend of Pluto's atmosphere, i.e. 
a monotonic increase of pressure since 1988.

\section{Central flash}
\label{sec_flash}

Nearly diametric occultation light-curves (but still avoiding the central flash) have flat bottoms (Fig.~\ref{fit_all}).
%
Our ray tracing code shows that near the shadow center, the stellar rays come
from a ``flash layer" about 3~km in thickness just above $r= 1191$~km, 
thus sitting 4~km on top of the assumed surface 
($R_P= 1187$~km, Fig.~\ref{fig_t_p_p_r}).
%

Let us denote by $F$ 
a model for the stellar flux (distinguishing it from the observed flux $F_\star$).
Deep inside Pluto's shadow, 
$F$ is roughly proportional to the local density scale-height, 
$H_n= -n/(dn/dr)= T/[\mu g/k +   (dT/dr)]$, where 
$\mu$ is the molecular weight, 
$g$ is the acceleration of gravity and
$k$ is Boltzmann's constant (DO15). 
For a spherical atmosphere, we have also $F \propto 1/z$, 
where $z$ is the distance to the shadow center.  
Writing $z= \sqrt{\rho^2+l^2}$, where 
$\rho$ is the closest approach distance to the shadow center and 
$l$ is the distance traveled from that point, we obtain:
\begin{equation}
F \propto \frac{H_n}{z} =
\frac{T}{\mu g/k +  dT/dr} \cdot
\left(\frac{1}{\sqrt{\rho^2+l^2}}\right).
\label{eq_flash}
\end{equation}
For an approximatively pure N$_2$ atmosphere ($\mu= 4.652 \times 10^{-26}$ kg), 
we obtain $\mu g/k \sim 2$~K~km$^{-1}$. 
As the thermal gradient $dT/dr$ is several degrees per kilometer at the flash layer (see below), 
the flash amplitude is significantly controlled by $dT/dr$.

Our best model 
minimizes 
$\chi^2= \sum_i 
\left\{
 \phi_i -
  \left[
   \left(1-\phi_P\right) F_i + \phi_P
  \right] 
\right\}^2/\sigma_i^2$,
where $\sigma_i^2$ is the variance of $\phi_i$ associated with the noise 
for the $i^{\rm th}$ data point.
As we do not measure $\phi_P$, 
we considered it as a free, adjustable parameter.
Among the data sets analyzed by DO15, only one had sufficient quality
- from  the 18 July 2012 ESO Very Large Telescope -
to permit a measurement of $\phi_P$
and thus constrain $dT/dr$ in the deepest accessible layer. 
It showed that the residual stellar flux, $F_{\rm res}$, at the bottom part of the light-curve
%
lay in the range 0.010-0.031,
thus imposing a thermal gradient near the surface
(and imposing $\phi_P$ for the other light-curves).
%
%
%
Since $F_{\rm res}$ was determined to within a factor of three,  
a large error bar on $dT/dr$ deep in Pluto's atmosphere was obtained, 
causing difficulties when extrapolating the pressure down to the surface.  
In doing so, we obtained a possible range $p_{\rm surf}=$~10-12~$\mu$bar for the surface pressure in 2012,
estimated at $r=1190 \pm 5$~km. 

As $F$ is roughly constant at the bottom of occultation light-curves 
(far from the flash), 
there is a degeneracy between $F$ and $\phi_P$: 
higher values of $\phi_P$ can be accommodated by smaller values of $F$,
i.e. smaller 
$H_n$.
This is \textit{not} true anymore within the flash, as $F$ suffers significant variations.
The $\chi^2$-minimization provides both $\phi_P$ and $H_n$ through 
$\partial \chi^2/\partial \phi_P= 0$ and
$\partial \chi^2/\partial H_n= 0$.
Although our ray tracing code generates exact values of $F$ for a given 
model, it is convenient here (for sake of illustration)  to note that $F$ is essentially proportional to 
$H_n$ (Eq.~\ref{eq_flash}), so that $\partial F/\partial H_n \sim F/H_n$.
Detailed calculations show that at minimum 
$\chi^2$, we have $\partial^2 \chi^2/\partial H_n^2 = (2N/H_n^2) (\sigma^2_F/\sigma^2)$ for $F \ll 1$, 
where $\sigma^2_F= \overline{F^2} - \overline{F}^2$ is the variance of $F$ 
(the bars denoting 
average values) and $N$ is the number of data  points. 
Thus, the relative error bar on the scale-height is $\delta H_n/H_n \sim (\sigma/\sigma_F)/\sqrt{N}$,
which is small if the flash (and then $\sigma_F$) is large.

Since $F$ increases as $H_n$ increases or $\rho$ decreases, 
$H_n$ and $\rho$ are correlated.
However, the full width at half maximum (FWHM) of the flash is proportional to $\rho$,
while $H_n$ controls homogeneously the flash amplitude, keeping its FWHM constant.
This disentangles the effects of $H_n$ and $\rho$.
More importantly, the BOOTES-3 and Dunedin stations exhibit 
flashes with similar amplitudes (Fig.~\ref{fit_all}).
This robustly forces the two stations to be
symmetrically placed with respect to the shadow center (Fig.~\ref{fig_chords}),
thus imposing $\rho \approx 45\pm2$~km for both stations, independently of $H_n$
(Table~\ref{tab_param}).
 
The $\chi^2$-value is minimized for   
$dT/dr= 8.5 \pm 0.25$~K~km$^{-1}$ at 1191~km in our model. 
This particular value must be considered with caution,
as it is not representative of the entire flash layer.
Due to the functional dependence of $T(r)$ (a branch of hyperbola, DO15),
the gradient $dT/dr$ varies rapidly around 1191~km. 
%
The average thermal gradient in the flash layer is in fact $\sim$5~K~km$^{-1}$, 
consistent with a previous flash analysis \citep{olk14}.
Besides, it is typical of what is expected from the heating by methane 
(D. Strobel 2015, private communication).
Other functional forms of $T(r)$ could be tested, but this remains outside the scope of this paper.
We note in passing that our best 2015 fit implies a residual stellar flux $F_{\rm res}=0.028$ 
(Fig.~\ref{fit_all}) that is compatible with the possible range (0.010-0.031) mentioned earlier for 2012.

Our spherical, transparent atmospheric model essentially captures the correct shape and height 
of the central flash (Fig.~\ref{fit_all}).
A closer examination of that figure reveals a small flux 
deficit (relative to the model) at the left side of the  BOOTES-3 flash.
It remains marginal, however, considering the general noise level. 
This said, it could be caused by an unmodeled departure 
of the flash layer from sphericity, but this is not anticipated. 
An atmosphere of radius $r$ rotating at angular velocity $\omega$ 
has an expected oblateness $\epsilon \sim r^3 \omega^2/2GM  \sim 10^{-4}$ 
for a rotation period of 6.4 days, $r \sim 1190$~km and 
Pluto's $GM$. 
Such oblateness causes a diamond-shaped caustic \citep{ell77} with a span 
of $4\epsilon r <\sim 1$~km  in the shadow plane. 
This is negligible considering the closest approach distances
involved  here ($\sim$45~km).
Moreover,  expected zonal winds of less than a few meters per second near 1191~km
\citep{van13,zal13} would have even smaller effects. 
More complex distortions may arise, as varying thermal conditions along Pluto's limb
may slightly tilt the local iso-density layer, but its modeling remains outside the scope of this paper.

A possible explanation of the small discrepancy  
is that the primary and/or secondary stellar images hit  
topographic features while 
moving around Pluto's limb. 
%
%
%
Curvature effects strongly stretch the images parallel to the limb during the central flash,
by a ratio equal to the flash layer radius (1191~km) divided by the
closest approach distance, about 45~km. 
From the star magnitudes (Table~\ref{tab_param} and \citealt{ker14}), 
we estimate its diameter as 33 $\mu$as, or 0.76~km projected at Pluto.
The length of the stellar image is then $0.76 \times (1191/45) \sim 20$~km.
It moves at about 4~km above the surface, which is comparable 
to the local topographic features reported from NH \citep{ste15}.
It is thus possible that part of the stellar flux was partially blocked by mountains, 
causing small observed drop. 
This can be tested by studying the topography derived from NH,  noting that
the primary and secondary stellar images at BOOTES-3 
probed regions near longitude 190$^\circ$E and latitude 20$^\circ$S, 
and 10$^\circ$E and 20$^\circ$N, respectively, during the flash.

Finally, NH images reveal tenuous hazes 
with normal optical depth $\tau_N \sim 0.004$ and scale-height $H=50$~km \citep{ste15}.
This implies an optical depth along the line of sight of 
$\tau \sim \sqrt{2\pi r/H} \cdot \tau_N \sim 0.05$, 
which is indistinguishable from the noise level (Fig.~\ref{fit_all}),
supporting our transparent-atmosphere hypothesis.

\section{Surface pressure}
 
Fig.~\ref{fig_t_p_p_r} displays our best pressure profile,
with $p_{1191}=11.0 \pm 0.2$~$\mu$bar at the deepest accessible level.
%
To estimate the surface pressure,  we need to extrapolate $p(r)$ into  the blind zone.
Two possible temperature profiles are considered, beside the DO15 model (Fig.~\ref{fig_t_p_p_r}).
One has a temperature gradient  in the blind zone that tends to zero at the surface,  
where  $p_{\rm surf}= 13.0$~$\mu$bar and $T_{\rm surf}=36$~K. \red
This describes a shallow troposphere that is in vapor pressure equilibrium with 
the surface, an example of a locally sublimating N$_2$ frost layer. \bla
The other profile has a constant gradient of 8.5~K~km$^{-1}$,
with $p_{\rm surf}= 12.6$~$\mu$bar and $T_{\rm surf}=49$~K. \red 
Such warmer regions are indeed observed on Pluto \citep{lel00}, and
they do not  sublimate due to the absence of free N$_2$ frost. 
\bla
Considering the formal error bar $\pm 0.2$~$\mu$bar on $p_{1191}$, 
\red
we obtain a range of 12.4-13.2~$\mu$bar for the surface pressure
under hypotheses (1)-(4) of Section~\ref{sec_evolution}, 
and 11.9-13.7~$\mu$bar accounting for the already dicussed possible bias of $\sim \pm0.5$~$\mu$bar.
\bla
%
Other thermal profiles should be considered at this point, 
but they would not change significantly our result
due to the proximity ($\sim$4~km) of our deepest accessible level to the surface,
leaving little freedom for $p_{\rm surf}$.

\section{Conclusions}

\begin{figure}
\epsscale{0.7}
\plotone{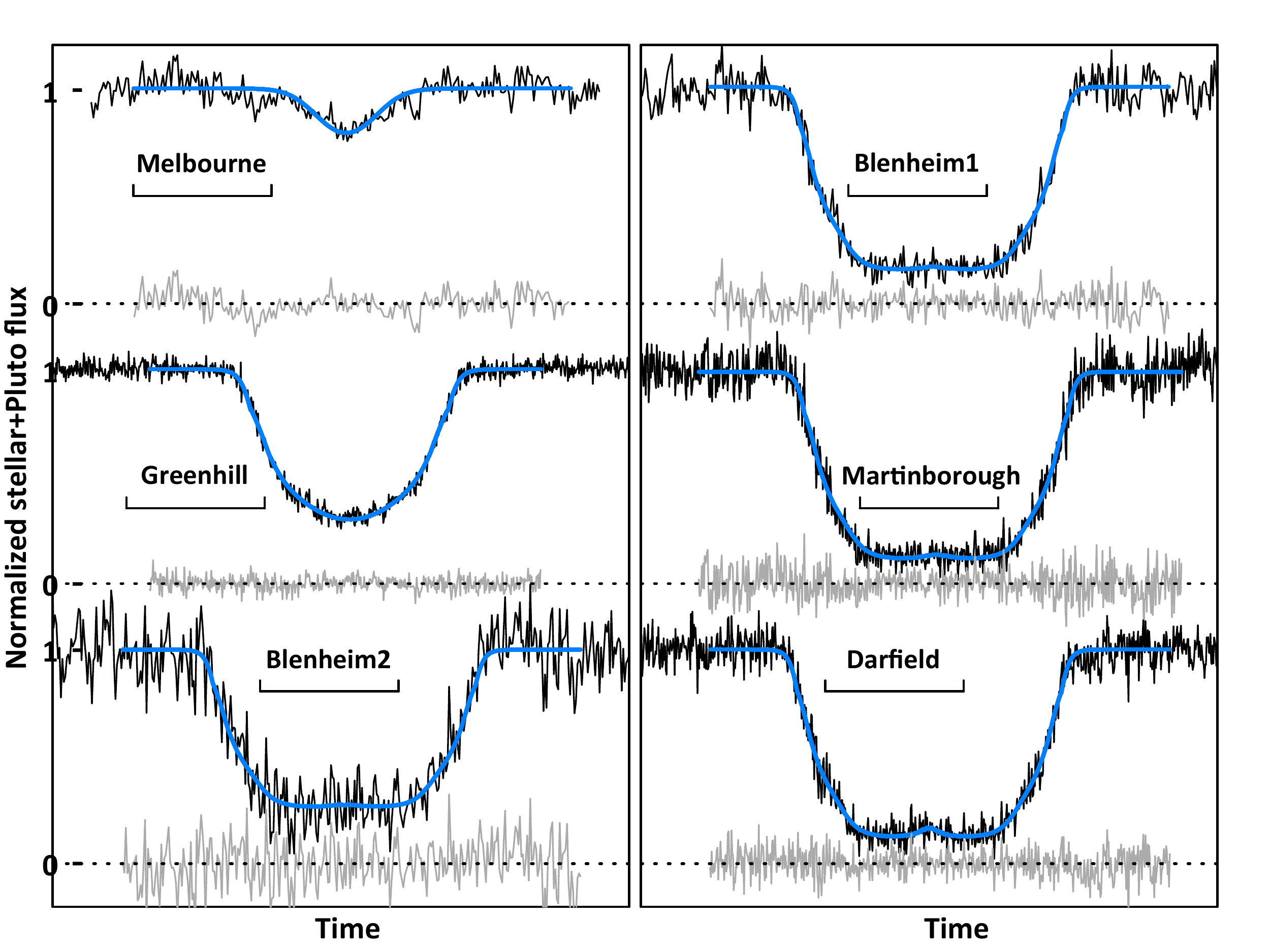}
\plotone{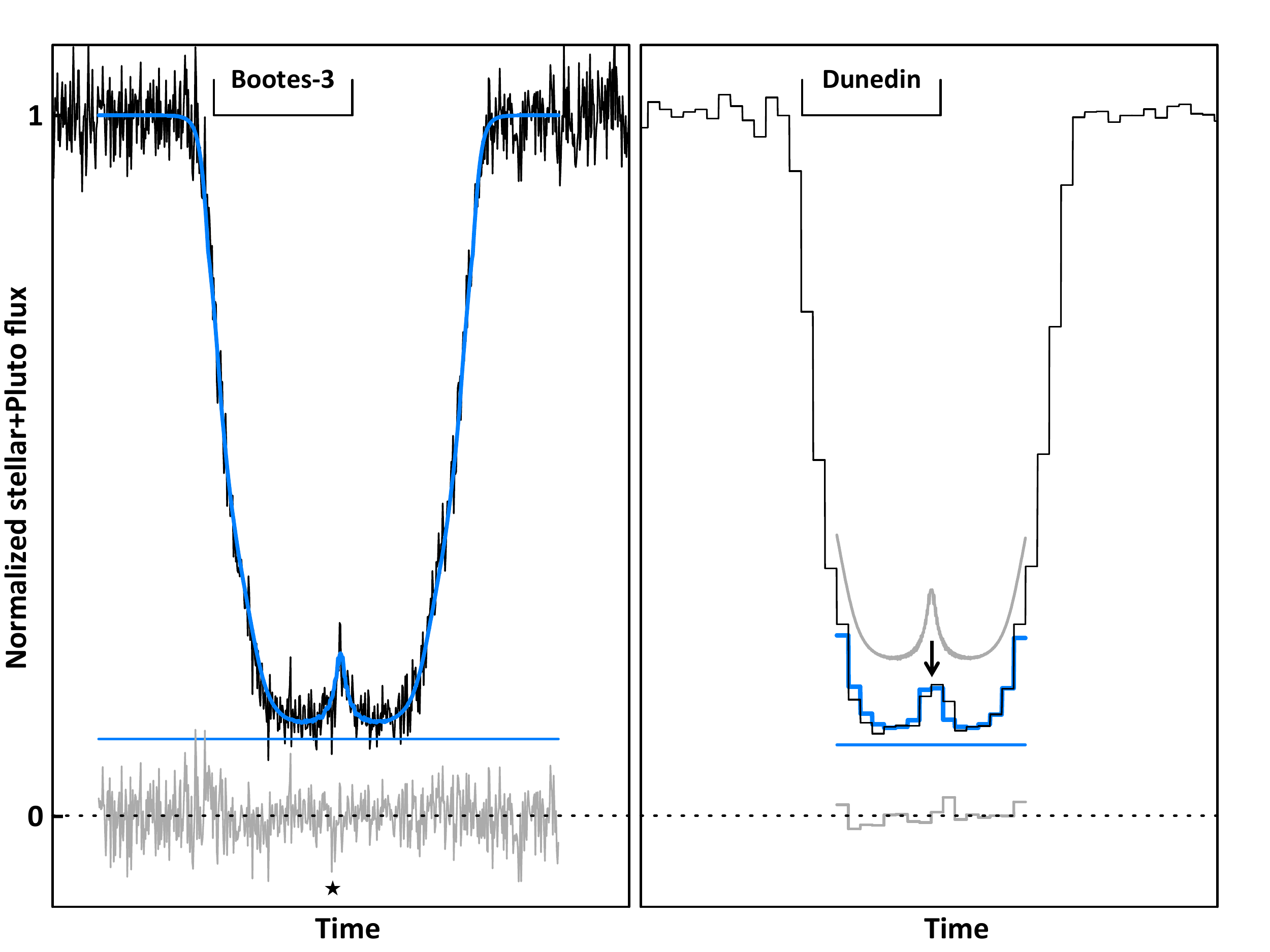}
\caption{%
Simultaneous fits to our 29 June 2015 occultation light-curves.
The intervals under each name correspond to the time-span
16h 52m-16h 43m UT.
The model is overplotted in blue, and  the residuals are in gray.
In the lower panels,  the blue horizontal lines are the fitted values of Pluto's contribution to the flux 
($\phi_P$, Eq.~\ref{eq_phi}).
The star symbol under the BOOTES-3 curve indicates a small flux deficit relative to the model.
In the Dunedin panel, the smooth curve is the central flash at high resolution,
before convolution by the exposure time (5.12~s),  
vertically shifted for better viewing.
\label{fit_all}
}
\end{figure}

The 29 June 2015 stellar occultation provided a snapshot of Pluto's atmosphere, 
after years of similar observations.
Moreover, this was the first event with multi-chord cuts into the central flash.
%
%
Assuming a spherical and transparent atmosphere as in DO15, 
we satisfactorily fit all the light-curves, including the central flash part (Fig.~\ref{fit_all}).

\begin{figure}
\epsscale{1.0}
\plotone{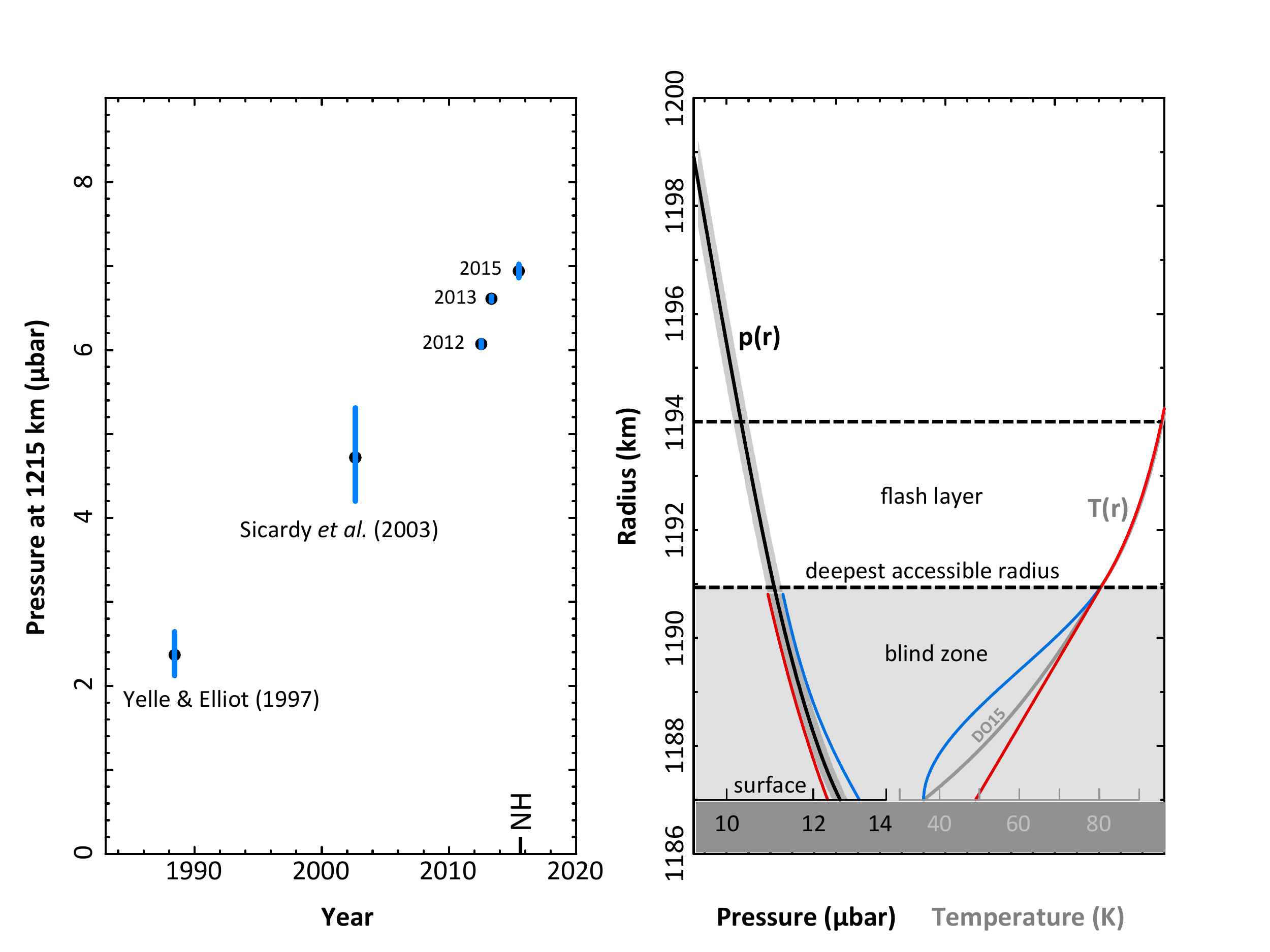}
\caption{%
Left: 
Pluto's atmospheric pressure at $r=1215$~km vs. time 
in 2012, 2013 and 2015 (our work), 
and from previous works \citep{yel97,sic03}, with 1$\sigma$ error bars.
%
The New Horizons Pluto flyby date (NH) is essentially coincident with our most recent dot.
Right:
our best pressure profile $p(r)$ for 29 June 2015, with formal 1$\sigma$-error domain.
The central flash layer roughly lies between the two horizontal dashed lines,
above the blind zone below 1191~km.
Two possible extrapolations (beside the DO15 model) of temperature profiles $T(r)$ into the blind zone are shown:
one with a thermal gradient that reaches zero at the surface (shallow troposphere, blue), and
one with a constant gradient 8.5~K~km$^{-1}$ (red).
%
\label{fig_t_p_p_r}
}%
\end{figure}

We find that Pluto's atmospheric pressure has been increasing monotonically since 1988, 
with an augmentation of $5 \pm 2$\% between 2013 and 2015, and 
an overall factor of almost three between 1988 and 2015 (Fig.~\ref{fig_t_p_p_r}). \red
This  trend between 1988 and 2013 was  confirmed by independent works by \cite{ell03,pas05,per13,you13,bos15}.
It is now extended to 2015 and \bla
rules out an ongoing atmospheric collapse associated with
Pluto's recession from the Sun. 
This is consistent with high thermal inertia models with a permanent N$_2$ ice cap 
over Pluto's north pole, that preclude such collapse \citep{olk15}.
Other possible models where N$_2$ condenses on an unlit cap might announce
a pressure decrease in the forthcoming years \citep{han15}.
Further monitoring 
with occultations and a detailed 
analysis of the NH data will allow discrimination between those scenarios.

The central flash comes from a $\sim$3-km-thick layer 
whose base is 4~km on top of Pluto's surface.
The amplitude of the flash is consistent with an average thermal gradient of $\sim$5~K~km$^{-1}$
in that layer.
Small departures from the model might be caused by
topographic features along Pluto's limb that block the stellar images.

Extrapolating possible temperature profiles down to the surface,
we find a possible range of 11.9-13.7~$\mu$bar for the surface pressure.
This is larger than, but compatible with the entry value $11 \pm 1$~$\mu$bar  
derived from the NH radio occultation experiment \citep{hin15,gla16}.
At this stage, more detailed investigations of both techniques should
be undertaken to see if this difference is significant, or the result 
of unaccounted effects.
%
In any case, the two techniques validate each other, an excellent prospect
for future monitoring of Pluto's atmosphere from ground-based occultations.

\acknowledgments

We acknowledge support from the French grant  ``Beyond Neptune II" ANR-11-IS56-0002,
and Labex ESEP.
The research leading to these results has received funding from
the European Research Council under the European Community's H2020
(2014-2020/ERC Grant Agreement 669416 ``LUCKY STAR").
EM acknowledges support  from the contrato de subvenci\'on 205-2014-Fondecyt, Peru.
JIBC  acknowledges the CNPq/PQ2 fellowship 308489/2013-6.
MA acknowledges the FAPERJ grant 111.488/2013,  CNPq/PQ2 fellowship 
312394/2014-4, and grants 482080/2009-4 and 473002/2013-2.
JLO acknowledges funding from Proyecto de Excelencia de la Junta de Andaluc\'{\i}a 
J.A.2012-FQM1776, Spanish grant AYA-2014-56637-C2-1-P, and FEDER funds.
AJCT acknowledges support from the Junta de Andaluc\'{\i}a (Project P07-TIC-03094) and 
Univ. of Auckland and NIWA for installing of the Spanish BOOTES-3 station 
in New Zealand, and support from 
the Spanish Ministry Projects AYA2012-39727-C03-01 and 2015-71718R
Development of the Greenhill Observatory was supported under the 
Australian Research Council's LIEF funding scheme (project LE110100055).
We thank C. Harlingten for the use of the H127 Telescope,
and D. and M. Warren for long term support.
We thank 
L. Beauvalet for running the ODIN Pluto's system model,
M.~W. Buie, S. Gwyn  and L.~A. Young  for providing pre-event Pluto's ephemeris and astrometry, 
D.~P. Hinson  and D.~F. Strobel for most useful discussions,
and the reviewer for useful comments 








\clearpage



\renewcommand{\baselinestretch}{1.45} 
\begin{table*}
\footnotesize
\renewcommand{\arraystretch}{0.65}   
\centering
\caption{Circumstances of observations.}
\vspace{0mm}
\label{tab_pos}
\begin{tabular}{lllll}
\hline
\hline
Site & Lat. (d:m:s)  & Telescope    &   Exp. Time/   & Observers \\
        & Lon. (d:m:s) & Instrument     &   Cycle (s)     & remarks \\
        & altitude (m)  &  Filter              &    & \\
\hline
\hline 
Melbourne         &   37 50 38.50  S  & 0.20 m            &   0.32   &  J. Milner    \\
Australia            & 145 14 24.40  E  & CCD/clear     &   0.32   &   occultation detected     \\
                           & 110                       &                         &             &     \\
\hline
Spring Hill       &   42 25 51.55 S     &   Harlingten/1.27 m &       0.1         &  A.~A. Cole,  A.~B. Giles    \\
Greenhill Obs.  & 147 17 15.49 E   &   EMCCD/B              &       0.1        &   K.~M. Hill     \\
Australia           & 650                        &                                     &                   & occultation detected         \\
\hline 
Blenheim1              &   41 32 08.59 S    &   0.28 m          & 0.64         &  G. McKay       \\
New Zealand         & 173 57 25.09 E    &   CCD/clear   &  0.64        &  occultation detected           \\
                                 & 18                           &                                   &       &                         \\
\hline
Blenheim2                 &    41 29 36.27 S    &  0.40 m          & 0.32      &  W.~H.  Allen   \\
New Zealand            &  173 50 20.72 E    &   CCD/clear  & 0.32      &    occultation detected         \\
                                    & 38                            &                         &               &                           \\
\hline
Martinborough       &    41 14 17.04 S    &  0.25 m      &    0.16   &    P. ~B. Graham   \\
New Zealand         &  175 29 01.18 E    &   CCD/B  &    0.16     &   occultation detected          \\
                                 & 73                            &                    &              &                            \\
\hline
Oxford                          &  43 18 36  S        &    0.35 m         &   1.28    &  S. Parker    \\
New Zealand              & 172 13 08  E      &    CCD/clear   &    1.28   &  occultation detected,  partially \\
                                      & 66 m                    &                          &              &  cloudy, not yet analyzed     \\
\hline
Darfield                    &    43 28 52.90 S    &  0.25 m           & 0.32         &    B. Loader        \\
New Zealand         & 172 06 24.04 E     &   CCD/clear    & 0.32         &    occultation detected, flash  \\
                                 & 210                      &                          &                     &                               \\
\hline
Christchurch           &   43 31 41 S    &  0.15 m           & 0.25         &    R. Glassey                  \\
New Zealand         & 172 34 54 E     &   CCD/clear    & 0.25         &    occultation detected  \\
                                 & 16                      &                          &                 &    not yet analyzed         \\
\hline
BOOTES-3 station  & 45 02 17.39 S         &  Yock-Allen/0.6m  & 0.34368       &   M. Jel\'{\i}nek       \\
Lauder                 & 169 41 00.88 E      &  EMCCD/clear    & 0.34463      &   occultation detected, flash  \\
New Zealand   	  &   370 	                 &                               &	                  &                              \\
\hline
Dunedin                &    45 54 31. S    &   0.35  m        & 5.12        & A. Pennell,  S. Todd    \\
New Zealand       & 170 28 46.  E    &   CCD/clear  &  5.12       &  M. Harnisch, R. Jansen       \\
                               & 118                    &                        &                 &  occultation detected, flash \\
\hline
\hline
Glenlee                &  23:16:09.6  S    & 0.30 m             & 0.32   &    S. Kerr      \\
Australia               & 150:30:00.8  E  & CCD/clear      & 0.32   &  no occultation detected          \\
                               & 50                       &                         &            &       \\
\hline
Reedy Creek        &  28 06 29.9  S    & 0.25 m           & 0.64  &  J. Broughton      \\
Australia                & 153 23 52.0  E  & CCD/clear     & 0.64   &  no occultation detected          \\
                               & 65                       &                          &                 &       \\
\hline
Linden                  &   33 42 30.0  S  & 0.76 m, 0.2 m & 0.133, 1.28    &    D. Gault,  R. Horvat   \\
Australia              &  150 29 43.5  E  & CCD/clear    &   0.133, 1.28   &    L. Davis    \\
                             &       583                &                        &                          &  no occultation detected     \\
\hline    
Leura                  &   33 43 09.0 S    & 0.20 m    &  n.a.      &    P. Nosworthy    \\
Australia            &  150 20 53.9  E  & visual     &  n.a.      &   no occultation detected    \\
                             &       903m            &                &              &       \\
\hline
Penrith               &   33 45 43.31  S   & 0.62 m           &   0.533  &  D. Giles     \\
Australia            &  150 44 30.30  E  &  CCD/Clear  &    0.533  &  M.~A. Barry    \\
                           & 96                       &                        &                   &  no occultation detected     \\
\hline
St Clair,              &   33 48 37  S        & 0.35 m           &   0.04   &  H. Pavlov     \\
Australia             & 150 46 37  E        & CCD/Clear   &   0.04   &   no occultation detected       \\
                            & 41                       &                       &                  &                        \\
\hline
Murrumbateman   &   34 57 31.50 S    & 0.40 m \& 0.35 m     & 0.16 \& 2   &  D. Herald, M. Streamer      \\
Australia                & 148 59 54.80 E    & CCD/clear               & 0.16 \& 2   &  no occultation detected   \\
                               & 594                    &                                        &            &       \\
\hline
Nagambie             &   36 47 05.71 S    &  0.20  m         & 0.64     & D. Hooper       \\
Australia                & 145 07 59.14 E    & CCD/clear    &   0.64   &  no occultation detected      \\
                               & 129                     &                       &                  &                       \\
\hline
\hline
\end{tabular}
\end{table*}
\normalsize 
\clearpage

\setlength{\tabcolsep}{2.1mm}		
\renewcommand{\baselinestretch}{1.28} 
\begin{table*}
\renewcommand{\arraystretch}{0.8}   
\centering
\caption{Input parameters and results}
\label{tab_param}
\begin{tabular}{llll}
\hline\hline
\multicolumn{4}{c}{Input parameters} \\ 
\hline 
\multicolumn{4}{c}{Star} \\ 
\hline 
Coordinates at epoch (J2000)\footnotemark[1]  & 
\multicolumn{3}{l}{$\alpha$= 19$^{\rm h}$ 00$^{\rm m}$ 49.4801$^{\rm s} \pm 11$~mas,  $\delta$= -20$^{\rm d}$ 41' 40.801"$\pm 17$~mas} \\ 
B, V, R, K magnitudes\footnotemark[2] & \multicolumn{3}{l}{12.8, 12.2, 12.8, 10.6} \\
\hline
\multicolumn{4}{c}{Pluto parameters} \\
\hline
\multicolumn{2}{l}{Pluto's geocentric distance, shadow velocity\footnotemark[3]} & \multicolumn{2}{l}{$4.77070 \times 10^9$~km, 24.1 km s$^{-1}$ (at 16:53 UT)} \\
\multicolumn{2}{l}{Pluto's mass and radius\footnotemark[4] \citep{ste15}}              & 
\multicolumn{2}{l}{$GM = 8.696 \times 10^{11}$ m$^3$ s$^{-2}$, $R_P = 1187$~km} \\
\multicolumn{2}{l}{Sub-observer and sub-solar latitudes\footnotemark[4]}             & \multicolumn{2}{l}{B= +51.66~deg, B'= +51.46~deg} \\
\multicolumn{2}{l}{Pluto's north pole position angle\footnotemark[4]}                      & \multicolumn{2}{l}{P=  +228.48~deg} \\
\hline
\hline
\multicolumn{4}{c}{Results} \\
\hline
\multicolumn{4}{c}{Thermal profile (input values for the DO15 model)} \\
\hline
$r_1$, $T_1$, $dT/dr(r_1)$, $r_2$, $T_2$  & \multicolumn{3}{l}{$1191.1$~km, 81.7 K, 8.5 K km$^{-1}$, $1217.3$ km, 109.7 K}  \\
$r_3$, $T_3$, $r_4$, $T_4$                          & \multicolumn{3}{l}{$1302.4$ km, 95.5 K, $1392.0$ km, 80.6 K}  \\
\hline
$c1,c2,c3$  & \multicolumn{3}{l}{$1.42143317 \times 10^{-3}$, $2.52794288 \times 10^{-3}$, $-2.12108557 \times 10^{-6}$} \\
$c4,c5,c6$  & \multicolumn{3}{l}{$-4.88273258 \times 10^{-7}$, $-7.04714651 \times 10^{-8}$, $-3.3716945  \times 10^{4}$} \\
$c7,c8,c9$  & \multicolumn{3}{l}{$7.7271133   \times 10^{1}$, $-5.86944930 \times 10^{-2}$, $1.48175559 \times 10^{-5}$} \\ 
\hline
\multicolumn{4}{c}{\red Longitudes and latitudes of half-light sub-occultation points\footnotemark[5]} \\
\hline
\multicolumn{4}{c}{\red ingress} \\
\hline
\multicolumn{4}{l}{\red %
Greenhill (154$^\circ$E, 06$^\circ$N, MT), 
Blenheim (120$^\circ$E, 28$^\circ$N, MT), 
Martinborough (119$^\circ$E, 28$^\circ$N, MT)}  \\
\multicolumn{4}{l}{\red %
Darfield (115$^\circ$E, 30$^\circ$N, MT), 
Bootes-3 (113$^\circ$E, 31$^\circ$N, MT), 
Dunedin (108$^\circ$E, 32$^\circ$N, MT)}  \\
\hline
\multicolumn{4}{c}{\red egress} \\
\hline
\multicolumn{4}{l}{\red %
Greenhill (232$^\circ$E, 37$^\circ$S, MT), 
Blenheim (280$^\circ$E, 35$^\circ$S, ET), 
Martinborough (282$^\circ$E, 34$^\circ$S, ET)}  \\
\multicolumn{4}{l}{\red %
Darfield (286$^\circ$E, 33$^\circ$S, ET), 
Bootes-3 (288$^\circ$E, 33$^\circ$S, ET), 
Dunedin (293$^\circ$E, 31$^\circ$S, ET)}  \\
\hline
\multicolumn{4}{c}{Pressure  \red (quoted errors at 1$\sigma$ level\footnotemark[6])} \\
\hline
 & 18 July 2012 & 04 May 2013 & 29 June 2015 \\
\hline
Pressure at 1215~km, $p_{1215}$ & $6.07\pm0.04$~$\mu$bar        & $6.61\pm0.03$~$\mu$bar & $6.94\pm \red 0.08$~$\mu$bar \\
Pressure at 1275~km, $p_{1275}$ & $2.09\pm0.015$~$\mu$bar      & $2.27\pm0.01$~$\mu$bar & $2.39\pm \red 0.03$~$\mu$bar \\
\multicolumn{3}{l}{Surface pressure (Fig.~\ref{fig_t_p_p_r})} & $11.9-13.7$~$\mu$bar \\
\hline
\multicolumn{4}{c}{Astrometry} \\
\hline
\multicolumn{2}{l}{Time of closest approach \red to shadow center (UT)} &  \multicolumn{2}{l}{Closest approach to shadow center}  \\
\hline
\multicolumn{2}{l}{BOOTES-3: 16$^{\rm h}$ 52$^{\rm m}$ $54.8\pm 0.1$$^{\rm s}$} &  \multicolumn{2}{l}{$45.9\pm 2$~km \red N of shadow center} \\
\multicolumn{2}{l}{Dunedin: 16$^{\rm h}$ 52$^{\rm m}$ $56.0\pm 0.1$$^{\rm s}$} &  \multicolumn{2}{l}{$44.6 \mp 2$~km \red S of shadow center}\\
\multicolumn{2}{l}{Geocenter: 16$^{\rm h}$ 55$^{\rm m}$  $04.9\pm 0.1$$^{\rm s}$} &  \multicolumn{2}{l}{$3911.5\pm2$~km \red N of shadow center} \\
\hline
\hline
\end{tabular}
~\\
\raggedright
\footnotemark[1]{See title's footnote for information.} \\
\footnotemark[2]{\cite{zac13,cut03,cut12}.} \\
\footnotemark[3]{PLU043/DE433 ephemeris.} \\
\footnotemark[4]{Using Pluto's north pole J2000 position: 
$\alpha_{\rm p}$= 08$^{\rm h}$ 52$^{\rm m}$ 12.94$^{\rm s}$, 
$\delta_{\rm p}$= -06$^{\rm d}$ 10' 04.8" \citep{tho08}.} \\
\footnotemark[5]{MT= morning terminator, ET= evening terminator.} \\
\footnotemark[6]{Formal errors. 
Possible systematic biases are $\pm 0.2,  \pm0.8$ and $\pm0.5$~$\mu$bar
in 2012, 2013 and 2015, respectively (Section~\ref{sec_evolution}).} \\
\end{table*} 
\renewcommand{\baselinestretch}{1.6} 

\clearpage

\end{document}